\documentclass{aa}  

\usepackage{graphicx}
\usepackage{txfonts}
\usepackage{graphicx}	
\usepackage{amsmath}	
\usepackage{amsfonts}
\usepackage{multicol}   
\usepackage{bm}		    
\usepackage{pdflscape}	
\usepackage{acro}
\usepackage[normalem]{ulem}
\usepackage{natbib}
\usepackage{hyperref}
\usepackage[capitalise]{cleveref}
\usepackage{widetext}
\usepackage{xcolor}
\usepackage{placeins}
\usepackage{afterpage}
\usepackage{longtable}

\newcommand{\iap}{CNRS \& Sorbonne Universit\'{e}, Institut d’Astrophysique de Paris (IAP), UMR 7095, 98 bis bd Arago, F-75014 Paris, France}
\newcommand{\vienna}{Heuristic and Evolutionary Algorithms Laboratory, University of Applied Sciences Upper Austria, Hagenberg, Austria}
\newcommand{\oxford}{Astrophysics, University of Oxford, Denys Wilkinson Building, Keble Road, Oxford OX1 3RH, UK}
\newcommand{\icg}{Institute of Cosmology \& Gravitation, University of Portsmouth, Dennis Sciama Building, Portsmouth, PO1 3FX, UK}
\newcommand{\cca}{Center for Computational Astrophysics, Flatiron Institute, 162 5th Avenue, New York, NY 10010, USA}
\newcommand{\Mpch}{\ensuremath{h^{-1}\text{Mpc}}}
\newcommand{\camb}{\textsc{camb}}
\newcommand{\colossus}{\textsc{colossus}}
\newcommand{\operon}{\textsc{operon}}
\newcommand{\fortran}{\textsc{fortran90}}
\newcommand{\python}{\textsc{python3}}
\newcommand{\bacco}{\textsc{bacco}}
\DeclareMathOperator{\aq}{aq}

\newcommand{\splitatcommas}[1]{%
  \begingroup
  \begingroup\lccode`~=`, \lowercase{\endgroup
    \edef~{\mathchar\the\mathcode`, \penalty0 \noexpand\hspace{0pt plus 1em}}%
  }\mathcode`,="8000 #1%
  \endgroup
}

\begin{document} 

   \title{A precise symbolic emulator of the linear matter power spectrum}

   \author{
    Deaglan J. Bartlett \thanks{\href{mailto:deaglan.bartlett@iap.fr}{deaglan.bartlett@iap.fr}} \inst{1}
    \and
    {Lukas Kammerer} \inst{2}
    \and
    {Gabriel Kronberger} \inst{2}
    \and
    {Harry Desmond} \inst{3}
    \and
    {Pedro G. Ferreira} \inst{4}
    \and
    {Benjamin D. Wandelt} \inst{1,5}
    \and
    {Bogdan Burlacu} \inst{2}
    \and
    {David Alonso} \inst{4}
    \and
    {Matteo Zennaro} \inst{4}
    }

   \institute{
        \iap
        \and
        \vienna
        \and
        \icg
        \and
        \oxford
        \and
        \cca
}

   \date{Received XXX; accepted YYY}

 
  \abstract
   {Computing the matter power spectrum, $P(k)$, as a function of cosmological parameters can be prohibitively slow in cosmological analyses, hence emulating this calculation is desirable. Previous analytic approximations are insufficiently accurate for modern applications, so black-box, uninterpretable emulators are often used.}
   {We aim to construct an efficient, differentiable, interpretable, symbolic emulator for the redshift zero linear matter power spectrum which achieves sub-percent level accuracy. We also wish to obtain a simple analytic expression to convert $A_{\rm s}$ to $\sigma_8$ given the other cosmological parameters.}
   {We utilise an efficient genetic programming based symbolic regression framework to explore the space of potential mathematical expressions which can approximate the power spectrum and $\sigma_8$. We learn the ratio between an existing low-accuracy fitting function for $P(k)$ and that obtained by solving the Boltzmann equations and thus still incorporate the physics which motivated this earlier approximation.
   }
   {We obtain an analytic approximation to the linear power spectrum with a root mean squared fractional error of 0.2\% between $k = 9\times10^{-3} - 9 \, h{\rm \, Mpc^{-1}}$ and across a wide range of cosmological parameters, and we provide physical interpretations for various terms in the expression. 
   Our analytic approximation is 950 times faster to evaluate than \camb{} and 36 times faster than the neural network based matter power spectrum emulator \bacco.
   We also provide a simple analytic approximation for $\sigma_8$ with a similar accuracy, with a root mean squared fractional error of just
   0.1\%
   when evaluated across the same range of cosmologies. This function is easily invertible to obtain $A_{\rm s}$ as a function of $\sigma_8$ and the other cosmological parameters, if preferred.  
   }
   {It is possible to obtain symbolic
   approximations to a seemingly complex function at a precision required for current and future cosmological analyses without resorting to deep-learning techniques, thus avoiding their black-box nature and large number of parameters. Our emulator will be usable long after the codes on which numerical approximations are built become outdated.
   }


   \maketitle
%

\section{Introduction}

Machine learning (ML) methods have great potential for
simplifying and accelerating the analysis of astrophysical data sets. 
The primary focus has been on what one might dub ``advanced numerical methods'' using, for example, Gaussian processes or neural networks. In these cases, one tries to construct efficient algorithms which can be used to either infer specific physical properties from complex data sets or emulate complex processes which can then be extrapolated to new situations. 
Typically these methods involve constructing a set of pre-established basis functions and then inferring their weights, or building complex, expressible functions, with parameters that can be optimised via efficient gradient descent methods. These methods can be easily incorporated in Bayesian inference frameworks that have achieved significant success, becoming the standard practice in astrostatistics.

The drawback of the more traditional, numerical ML techniques is their opaqueness; it is not always clear what information is being used and how methods trained on (necessarily imperfect) simulations will perform when applied to real-world data. 
A somewhat overlooked branch of machine learning which has tremendous promise for the types of problems being considered in astrophysics is symbolic regression (SR). 
With SR one tries to infer the mathematical expressions that best capture the properties of the physical system one is trying to study. The process is an attempt to mimic and systematise the practice that physicists have always used: to infer simple physical laws (i.e. formulae) from data.
The field of SR has developed over the years into a vibrant and active field of research in ML, typically associated with evolutionary methods such as Genetic Programming. 
It has been shown that it can be used to infer some well-established laws of physics from data and infer new ones \citep[e.g.][]{Lemos_2022,Bartlett_2022,Bartlett_2023,Desmond_2023,Kamerkar_2022,Sousa_2023,Delgado_2022,Miniati_2022,Wadekar_2022,Koksbang_2023b,Koksbang_2023c,Koksbang_2023a,Alestas_2022,Lodha_2023}.

Within the field of cosmology, one often compresses observations from galaxy surveys into two-point correlation functions (or their Fourier transforms, power spectra), which are compared to theory through Markov Chain Monte Carlo methods to constrain cosmological parameters. As cosmological surveys become increasingly vast and precise, a fundamental limitation to the feasibility of such inferences has been the speed at which one can make this theoretical prediction, since it involves solving a complex set of coupled, highly nonlinear differential equations. 
Recently, instead of directly solving these equations \citep{Lewis_2000,Blas_2011,Hahn_2023} and adding non-linear corrections \citep{Smith_2003}, emulation techniques such as Neural Networks, Gaussian Processes or polynomial interpolation schemes have been used to accelerate these calculations to directly output the matter power spectrum as a function of cosmological parameters \citep{Pico2,Pico1,Heitmann_2014,Winther_2019,Angulo_2021,Arico_2021,Knabenhans_2021,SpurioMancini_2022,Mootoovaloo_2022,Zennaro_2023}.
These methods act as black boxes and require up to several hundreds of parameters to be optimised. 

However, through perturbation theory, one knows analytic limits of the power spectrum and, through visual inspection, it does not appear to be an extremely complex function. 
As such, one wonders whether an analytic approximation exists. Indeed, for many years, the leading method of accelerating this calculation has been an analytic approximation \citep{Eisenstein_1998,Eisenstein_1999}, however it is insufficiently precise for modern experiments.
Analytic approximations to beyond $\Lambda$CDM power spectra have been proposed in the context of modified gravity \citep{Bayron-Orjuela-Quintana_2023}, although these still only achieve a precision of between 1 and 2\%.

Such an emulator has the advantage that it will not become deprecated when the codes on which current numerical methods are built become outdated, whereas other methods require the transfer of the inferred weights and biases as well as the model architecture, hindering longevity. Even in the short term, an analytic expression using standard operators is more portable, since it can be more easily be incorporated into the user's favourite programming language without the need to install or write wrappers for the model. Moreover, having an analytic expression allows one to interpret such a fit, and potentially identify physical processes which could lead to certain terms, contrary to the black-box numerical methods. Additionally, such expressions often contain fewer free parameters to optimise than numerical ML methods.

In \cref{sec:Pofk} we briefly describe the matter power spectrum and the Eisenstein \& Hu approximation, and in \cref{sec:SR} we detail the SR method we use in this work. We present an analytic emulator for $\sigma_8$ as a function of other cosmological parameters in \cref{sec:sigma8_emulator} (which is easily invertible to obtain $A_{\rm s}$ as a function of cosmological parameters), and in \cref{sec:pofk_emulator} we give our emulator for the linear matter power spectrum. The main results of this paper are given in \cref{eq:sigma8_fit} and \cref{eq:pk_lin_fit}. We conclude and discuss future work in \cref{sec:Conclusion}.
Throughout this paper `$\log$' denotes the natural logarithm.

\section{The matter power spectrum}
\label{sec:Pofk}

\subsection{Definition}

We would like to construct an efficient, differentiable and (if at all possible) interpretable emulator for the  power spectrum of the matter distribution in the Universe, $P(k; \bm{\theta})$, for wavenumber $k$ and cosmological parameters $\bm{\theta}$. 

The power spectrum is defined as follows: the matter density of the Universe, $\rho(\bm{x})$ can be decomposed into a constant (in space) background density, ${\bar \rho}$, and a density contrast, $\delta(\bm{x})$ such that $\rho(\bm{x})={\bar\rho}[1+\delta(\bm{x})]$. If 
$\tilde{\delta}(\bm{k})$ is the Fourier Transform of $\delta(\bm{x})$, and the matter distribution is statistically homogeneous and isotropic, we have that
\begin{equation}
    (2 \pi)^3 P(k;\bm{\theta}) \delta^{\rm D} \left( \bm{k} - \bm{k}^\prime \right)\equiv\langle \tilde{\delta}(\bm{k}) \tilde{\delta}^\ast (\bm{k}^\prime) \rangle,
\end{equation}
where $\langle\cdots\rangle$ denotes an ensemble average and $\delta^{\rm D}$ is the Dirac delta function.

From observations of the Cosmic Microwave Background (CMB) \citep{Planck_VI_2018}, it is known that the density fluctuations at early times were approximately Gaussian and thus fully described by $P(k; \bm{\theta})$. At these early times, the power spectrum of the comoving curvature perturbations is proportional to $A_{\rm s}k^{n_{\rm s} - 4}$, where $n_{\rm s} \approx 0.9665$ \citep{Planck_VI_2018}. Although structure formation through gravity makes the present-day density field non-Gaussian (e.g. the intricate structure of the cosmic web is typically associated with higher order statistics), the power spectrum still holds a central role in modern cosmological analyses.

The current cosmological model is described by only six parameters: the baryonic, $\Omega_{\rm b}$, and total matter, $\Omega_{\rm m}$, density parameters, the Hubble constant, $H_0 = 100 h {\rm \, km \, s^{-1} \, Mpc^{-1}}$, the scalar spectral index, $n_{\rm s}$, the curvature fluctuation amplitude, $A_{\rm s}$, and the reionisation optical depth, $\tau$. All other parameters can be derived from these six, and thus sometimes a different set of parameters is chosen. For example, instead of $A_{\rm s}$, one often quotes $\sigma_8$ which is 
the root-mean-square density fluctuation when the linearly evolved field is smoothed with a top-hat filter of radius $8 \,\Mpch$. Specifically, one defines for a top-hat of radius $R$
\begin{equation}
    \label{eq:sigmaR}
    \sigma_R^2 = \int {\rm d}k \, \frac{k^2}{2 \pi^2} P (k; \bm{\theta}) \left| W(k,R) \right|^2,
\end{equation}
where $\bm{\theta}$ is the set of cosmological parameters and the Fourier transfer of the top-hat filter is 
\begin{equation}
    W(k, R) = \frac{3}{(kR)^3} \left( \sin (k R) - kR \cos (k R) \right),
\end{equation}
and $\sigma_8$ is simply $\sigma_R$ for $R = 8 \Mpch$.
Throughout this paper we ignore the small dependence of the power spectrum on the reionisation optical depth parameter, and focus on the remaining five parameters. We set the neutrino mass to zero in all calculations.

\subsection{Eisenstein \& Hu approximation}

Since each evaluation of a Boltzmann solver to compute $P(k; \bm{\theta})$ can be expensive, the ability to emulate this procedure and replace this solver with a surrogate model has long been desirable. The most notable attempt to do this in an analytic manner is given in a series of papers by \citet{Eisenstein_1998,Eisenstein_1999}. In these works, an approximation is constructed based on physical arguments including baryonic acoustic oscillations (BAO), Compton drag, velocity overshoot, baryon infall, adiabatic damping, Silk damping, and cold dark matter growth suppression.
Rather than repeat their findings, we refer the reader to these papers to inspect the structure of the equations.
Such a model is accurate to a few percent which, although invaluable at the time of writing, is insufficiently accurate for modern cosmological analyses. It is thus the goal of this work to build upon this analytic emulator to provide sub-percent level predictions.
We note that alternative symbolic approximations also exist to $P(k)$, such as the earlier, less accurate approximation by \citet{Bardeen_1986} (BBKS). More recently, \citet{Bayron-Orjuela-Quintana_2022} found simple expressions using genetic programming which can achieve similar accuracy to the \citeauthor{Eisenstein_1998} expression, but we choose to use \citeauthor{Eisenstein_1998}'s approximation due to its physical motivation and widespread use.

\section{Symbolic regression}
\label{sec:SR}

To extract analytic approximations from sampled data, we use the symbolic regression package \operon{}\footnote{\url{https://github.com/heal-research/operon}} \citep{Burlacu_2020}. 
This package leverages the most popular 
\citep[e.g.][]{Lemos_2022,cranmer2020discovering,pysr,Cranmer_2023,Schmidt_2009,Schmidt_2011,Virgolin_2019,deFranca_2021,LaCava_2019,Kommenda_2020,Virgolin_2021,Arnaldo_2014}
approach to SR, namely genetic programming \citep{turing,David, haupt}.
Genetic programming describes the evolution of ``computer programs'', in our case mathematical expressions encoded as expression trees. Following the principle of natural selection, over several iterations the worst performing equations (given some fitness metric) are discarded and new equations are produced by combining sub-expressions of the current population (crossover) or by randomly inserting, replacing or deleting a subtree in an expression (mutation). Over the course of several generations, the expectation is that the population of equations evolve to become fitter and thus we obtain increasingly accurate analytic expressions. 

We note that many other techniques exist for SR, such as 
supervised or reinforcement learning with neural networks \citep{Petersen_2021,Landajuela_2022,Tenachi_2023,Biggio_2021}, 
deterministic approaches \citep{Worm, Kammerer_2021, Dome,FFX}, 
Markov chain Monte Carlo \citep{Jin_2019}, 
physics-inspired searches \citep{aifeyn_0,aifeyn,QLattice},
and exhaustive searches \citep{Bartlett_2022}. However, we choose \operon{} and thus genetic programming due to its speed, high memory efficiency and its strong performance in benchmark studies \citep{LaCava_2021,Burlacu_2023}.

To improve the search, every time a terminal node appears in an expression tree (i.e. $k$ or one of the cosmological parameters), a scaling parameter is introduced, which is then optimised~\citep{Kommenda_2020} using the Levenberg–Marquardt algorithm \citep{Levenberg_1944,Marquardt_1963}. We denote the total number of nodes in the expression excluding the scaling as the ``length'' of the model, and the ``complexity'' refers to the total number of nodes, including these.

When comparing objective values during non-dominated sorting (NSGA2), \operon{} implements the concept of $\epsilon$-dominance \citep{Laumanns_2002}, where the parameter $\epsilon$ is defined such that two objective values which are within $\epsilon$ of each other are considered equal. This parameter therefore affects the number of duplicate equations in the population and is designed to promote convergence to a representative well distributed approximation of the global Pareto front:
the set of solutions which cannot be made more accurate without being made more complex.
We choose different values for this parameter when searching for our two emulators, and these were found after some experimentation with different values to find settings which produced accurate yet compact models.

Model selection is an essential part of any SR search. Since one optimises both accuracy and simplicity during the search, SR is often a Pareto-optimisation problem. 
In the presence of statistical errors, one can combine simplicity and accuracy in a principled, information theory motivated way into a single objective to optimise under the minimum description length principle \citep{Bartlett_2022}. In this case, the task of picking the optimum function is unambiguous, and one can incorporate prior information into the functional form using language models \citep{Bartlett_2023} to obtain more physically motivated functions.
In our problem, however, we do not have noise in our data and thus have to rely on heuristic methods.

To choose a model, we first generate the Pareto front of candidate expressions. We then consider only those models for which the loss is below some predefined level (to ensure sufficiently accurate solutions for our applications) and those for which the loss on the training and validation sets do not differ significantly (and indicator of over-fitting).
At this point one could automate model selection; for example, in the code \textsc{pysr} \citep{cranmer2020discovering,pysr} the best model is the one with the best ``score'', which is the one with largest negative of the derivative of the loss with respect to complexity. However, given we wish to have interpretable and physically-reasonable functions, we instead visually inspect the most accurate solution found for each model length and make a qualitative judgement as to the function which is sufficiently compact to be interpretable yet is accurate enough for our applications. 

Further details are given in \cref{sec:sigma8_emulator,sec:pofk_emulator}.

\section{Analytic emulator for \texorpdfstring{$\sigma_8$}{sigma8}}
\label{sec:sigma8_emulator}

\begin{table}
    \caption{Cosmological parameters used for analytic emulators.}
    \centering
    \begin{tabular}{c|c|c}
        Parameter & Minimum & Maximum \\
        \hline\hline
         $10^9 \, A_{\rm s}$ & 1.7 & 2.5\\
         $\Omega_{\rm m}$ & 0.24 & 0.40 \\
         $\Omega_{\rm b}$ & 0.04 & 0.06 \\
         $h$ & 0.61 & 0.73 \\
         $n_{\rm s}$ & 0.92 & 1.00 \\
    \end{tabular}
    \tablefoot{We sample all parameters independently and uniformly in the range between the minimum and maximum values given.}
    \label{tab:cosmo_par_prior}
\end{table}

\begin{figure}
    \centering
    \includegraphics[width=0.65\columnwidth,height=0.29\columnwidth]{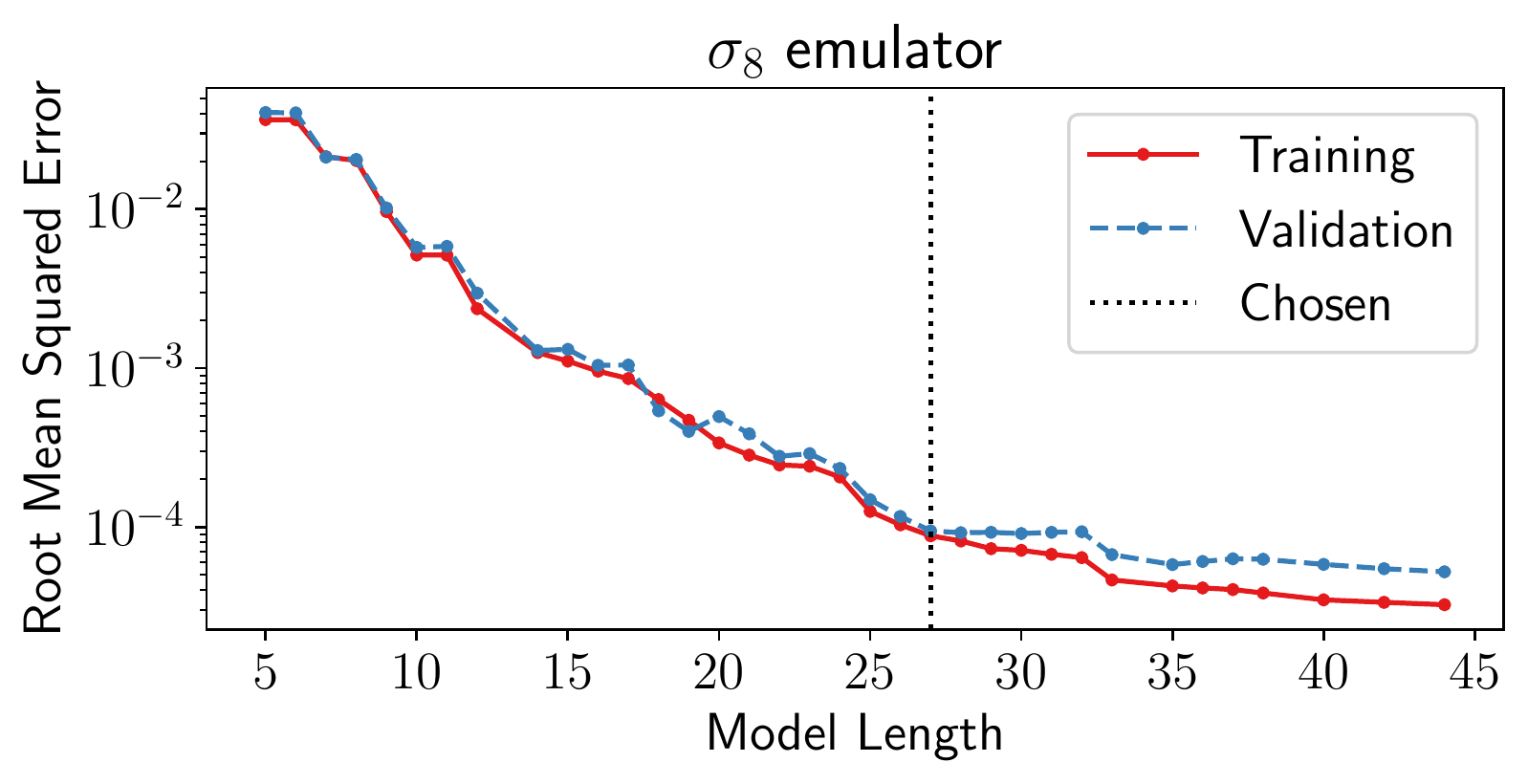}
    \caption{Pareto front of solutions obtained using \operon{} when fitting 
    $\sigma_8 / \sqrt{10^9 A_{\rm s}}$
    as a function of 
    $\Omega_{\rm b}$, $\Omega_{\rm m}$, $h$ and $n_{\rm s}$. 
    We plot the root mean squared error as a function of model length from the training and validation sets separately. The model in \cref{eq:sigma8_fit} has a model length of 27.
    }
    \label{fig:sigma8_pareto}
\end{figure}

We begin by considering the simplest emulator one may want for power spectrum related quantities: an emulator for $\sigma_8$ as a function of other cosmological parameters ($A_{\rm s},\Omega_{\rm b},\Omega_{\rm m}, h,  n_{\rm s}$) or, equivalently, an emulator for $A_{\rm s}$ given $\sigma_8$ and the other cosmological parameters. 
Although the set of neural network emulators \bacco{} contains a function to do this \citep{Arico_2021}, to the best of the authors' knowledge an analytic approximation is not currently in common use.
The standard approach is to compute the linear matter power spectrum with a Boltzmann code assuming some initial guess of $A_{\rm s}$, then compute the integral in \cref{eq:sigmaR} to obtain $\sigma_8$. For a target $\sigma_8$ of $\sigma_8^\prime$, one should then use $A_{\rm s}^\prime = (\sigma_8^\prime/\sigma_8)^2 A_{\rm s}$. 

We wish to accelerate this process with a symbolic emulator.
To compute this, we constructed a Latin hypercube (LH) of 100 sets of cosmological parameters, using uniform priors in the ranges given in \cref{tab:cosmo_par_prior}, which are the same as those used in \citet{Knabenhans_2021}. 
We constructed a second LH of 100 points to be used for validation. For these parameters, we computed $\sigma_8$ using \camb{} \citep{Lewis_2000} and attempted to learn this mapping using a mean squared error loss function with \operon. 

For the equation search, we used a population size of 1000 with a brood size of 10 and tournament size of 5, optimising both the mean squared error and the length of the expression simultaneously, with 
$\epsilon = 10^{-6}$
(see \cref{sec:SR}). 
From \cref{eq:sigmaR}, one would expect that $A_{\rm s}$ only appears in the expression for $\sigma_8$ as $\sigma_8 \propto \sqrt{A_{\rm s}}$ since $A_{\rm s}$ linearly scales the power spectrum. As such, we chose to fit for $\sigma_8 / \sqrt{10^9 A_{\rm s}}$, where we use $10^9 A_{\rm s}$
instead of $A_{\rm s}$ so that all cosmological parameters and the target variable are $\mathcal{O}(1)$.
Parameters were optimised during the search using a nonlinear least squared optimiser with up to 1000 iterations per optimisation attempt. We set the maximum allowed model length to 
40
and maximum number of iterations to 
$10^8$,
although we found that both of these are much larger than the required values needed to converge to a desirable solution.

The candidate expressions were comprised of standard arithmetic operations (addition, subtraction, multiplication),  as well as the natural logarithm, square and square root operators.
It is somewhat difficult to predict the exact effect the function set has on the quality of the results. The efficiency of the algorithm is only affected insofar as the transcendental functions ($\sin$, $\cos$, $\log$, $\exp$, $\sqrt{\cdot}$, etc.) are slower to evaluate than arithmetic operators. The effect is minor, however. Increasing the number of basis functions inflates the total search space, making it potentially less likely that well-fitting expressions are found, yet too small a basis set could remove compact approximations to the functions of interest. For both this section and \cref{sec:pofk_emulator} we experimented with alternative basis sets and found that those chosen gave accurate yet compact expressions within a reasonable run time.

After
2
minutes of operation on one node of 
56
cores, 
we found the Pareto front of expressions given in \cref{fig:sigma8_pareto}. We see that the training and validation losses are comparable at all model lengths, reaching a root mean squared error of around
$10^{-3}$ by a model length of 14. We see that the difference between the training and validation losses increases after the model of length 27, so we take this model as our fiducial result. It is given by
\begin{equation}
    \begin{split}
    \label{eq:sigma8_fit}
    \frac{\sigma_8}{\sqrt{10^9 A_{\rm s}}} &\approx
    a_0 \Omega_{\rm m} + a_1 h + a_2 
    \left( \Omega_{\rm m} - a_3 \Omega_{\rm b} \right) \left( \log (a_4 \Omega_{\rm m}) - a_5 n_{\rm s} \right) \\
    & \times \left( n_{\rm s} + a_6 h \left( a_7 \Omega_{\rm b} - a_8 n_{\rm s} + \log (a_9 h) \right)\right),
     \end{split}
\end{equation}
where the optimised parameters are 
$\bm{a}= \splitatcommas{[0.51172, 0.04593, 0.73983, 1.56738, 1.16846, 0.59348, 0.19994, 25.09218, 9.36909, 0.00011]}$.
We note that we have removed a final additive term produced by \operon{} since this has a value of $3 \times 10^{-6}$ and is thus much smaller than the error in the fit so can be safely neglected.

We note several important features of this equation which make it desirable. First, we find that it is a highly accurate approximation, with a root mean squared fractional error on the validation set of only 
0.1\%
which is far smaller than the precision to which one can measure this number with cosmological experiments.
Second, one sees that $A_{\rm s}$ 
(by design)
only appears once in this equation and as a 
multiplicative
term. 
Thus, it is trivial to invert this equation to obtain $A_{\rm s}$ as a function of the other cosmological parameters, as is often needed. 

\section{Analytic emulator for the linear power spectrum}
\label{sec:pofk_emulator}

\begin{figure}[htb]
    \centering
    \includegraphics[width=\columnwidth]{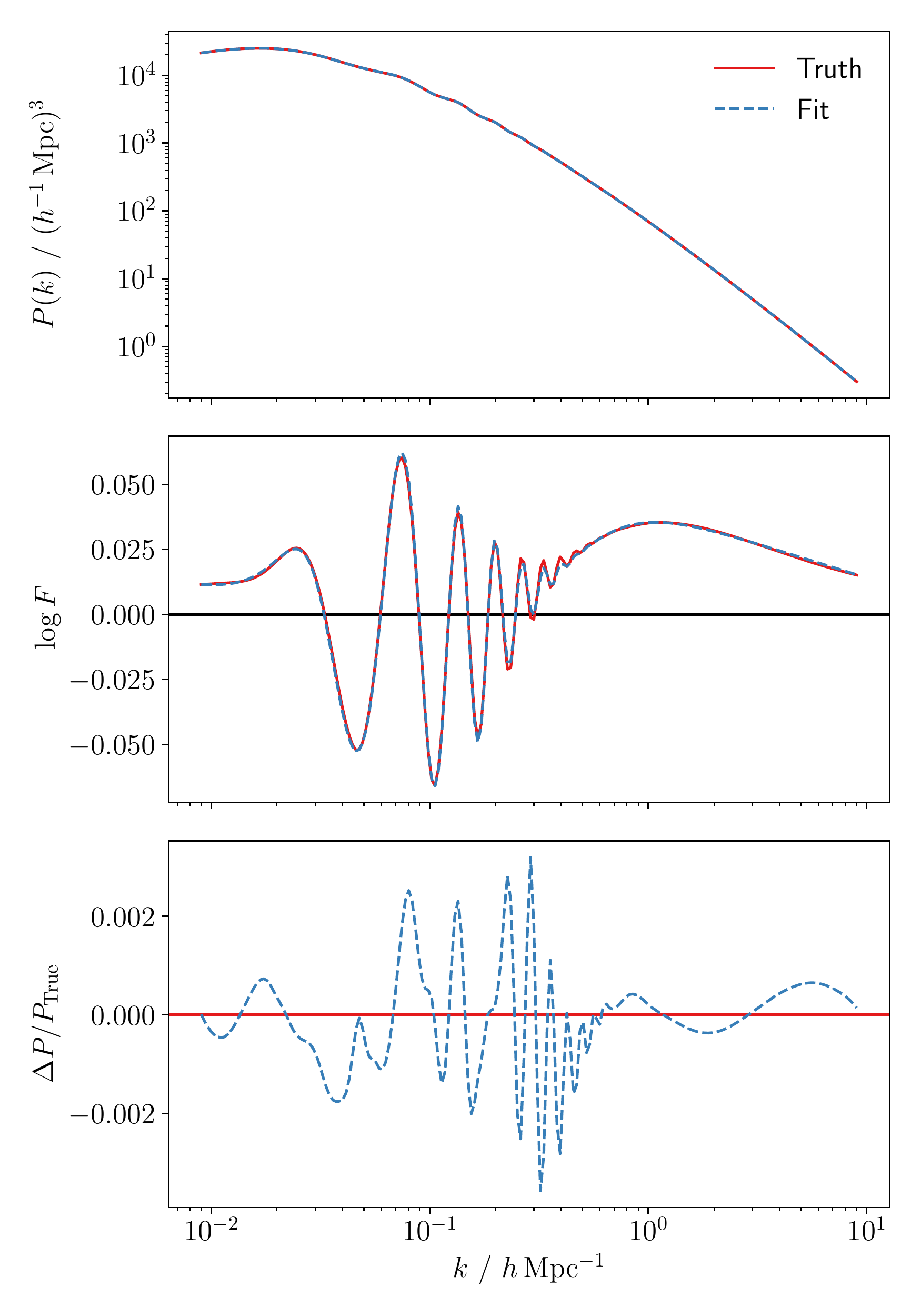}
    \caption{Linear matter power spectrum (upper), the residuals (\cref{eq:Pk_residual_definition}) from the \citeauthor{Eisenstein_1998} fit without baryons (middle), and the fractional residuals on $P(k)$ compared to the truth for the Planck 2018 \citep{Planck_VI_2018} cosmology. In all panels we plot the truth computed with \camb{} with solid red lines, and the analytic fit (\cref{eq:pk_lin_fit}) obtained in this paper with dashed blue lines. We see that the fit is accurate within 0.3\% across all $k$ considered. 
    }
    \label{fig:planck_fit_linear}
\end{figure}

We now move on to the more challenging task of producing an analytic emulator for the linear matter power spectrum. Given the previous success of \citet{Eisenstein_1998,Eisenstein_1999}, we believe it is sensible to build upon this work, not least due to the physically-motivated terms included in their fit and so that we must only have to fit a small residual (of the order of a few percent). Thus, instead of directly fitting for $P(k, \bm{\theta})$, we define
\begin{equation}
    \label{eq:Pk_residual_definition}
    P(k; \bm{\theta}) \equiv P_{\rm EH}(k; \bm{\theta}) F(k; \bm{\theta}),
\end{equation}
where $P_{\rm EH}(k; \bm{\theta})$ is the zero-baryon fit of \citet{Eisenstein_1998}, which does not include an attempt to fit the BAO. We plot both $P(k; \bm{\theta})$ and $\log F(k; \bm{\theta})$ in \cref{fig:planck_fit_linear} for the best-fit cosmology obtained by Planck \citep{Planck_VI_2018}, where we see that dividing out the \citeauthor{Eisenstein_1998} term retains the BAO part of the power spectrum and reduces the dynamic range required for the fit.

As before, we obtained 100 sets of cosmological parameters on a LH using the priors in \cref{tab:cosmo_par_prior} and computed both $P(k;\bm{\theta})$ with \camb{} and $P_{\rm EH}(k; \bm{\theta})$ with the \colossus{} \citep{Diemer_2018} implementation, using 200 logarithmically spaced values of $k$ in the range $9\times 10^{-3} - 9 \, h {\rm \, Mpc^{-1}}$. We note that this is an extremely small training set compared to many power spectrum emulators, but we find that it is sufficient to obtain sub-percent level fits.

We chose to symbolically regress $\log F (k; \bm{\theta})$ using a mean squared error loss function, and thus wish to minimise the fractional error on this residual. We chose to fit for $\log F$ as this ensures that our final estimate of $P(k; \bm{\theta})$ is positive, as guaranteed by exponentiation, which is physically required. Additionally, we first multiplied $\log F$ by 100 so that the target was $\mathcal{O}(1)$. We used a further 100 sets of cosmological parameters, also arranged on a LH, for validation. We chose to fit using the cosmological parameters $\sigma_8$, $\Omega_{\rm b}$, $\Omega_{\rm m}$, $h$ and $n_{\rm s}$. We used the same settings for \operon{} as in \cref{sec:sigma8_emulator}, except we chose $\epsilon=10^{-3}$,  terminated our search after $10^8$ function evaluations and used a basis set comprising of addition, subtraction, multiplication, natural logarithm, cosine, power and analytic quotient operators ($\aq(x,y) \equiv x / \sqrt{1+y^2}$).

\begin{figure}[htb]
    \centering
    \includegraphics[width=0.95\columnwidth]{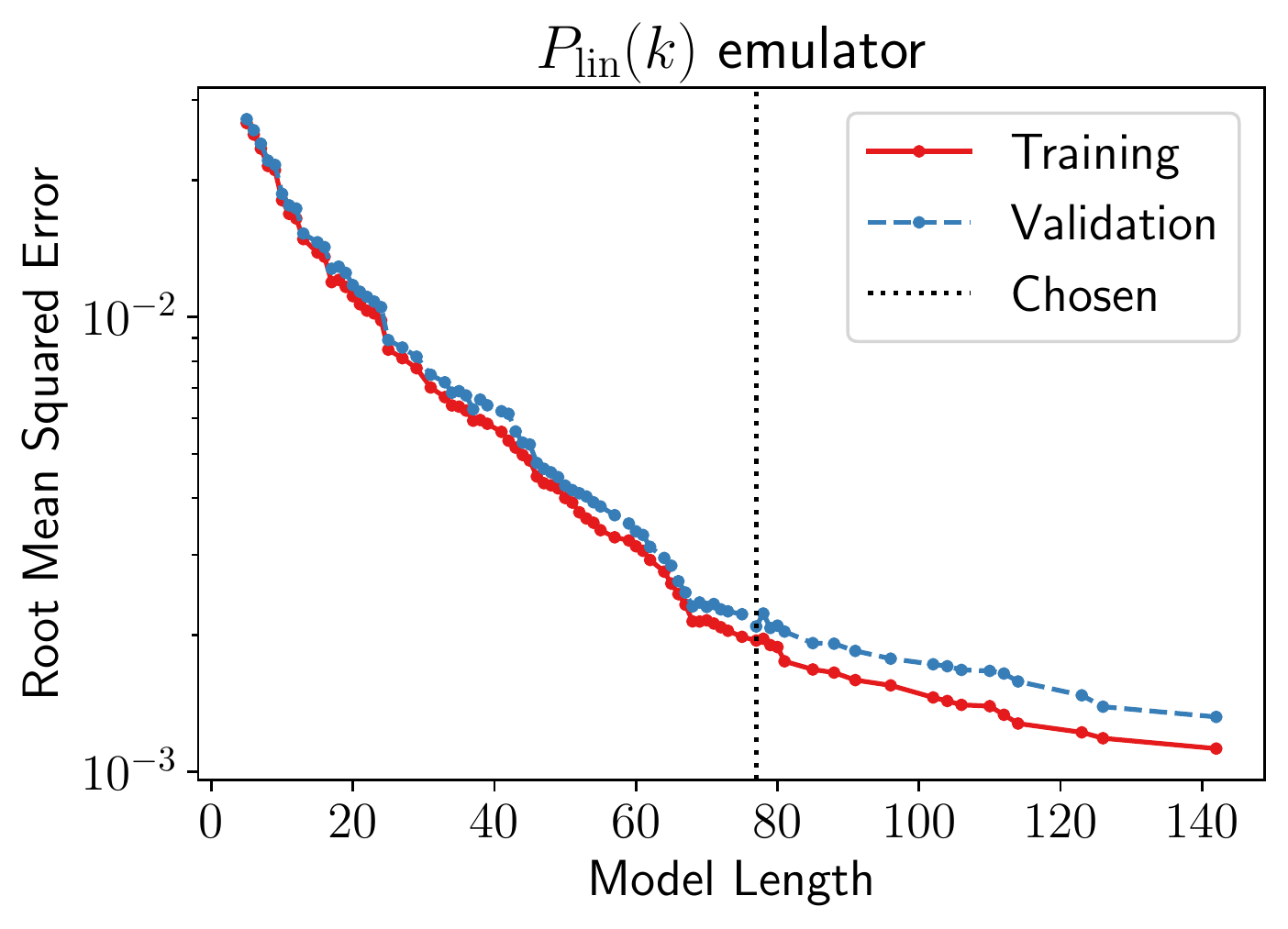}
    \caption{Pareto front of solutions obtained using \operon{} when fitting the linear matter power spectrum as a function of $\sigma_8$, $\Omega_{\rm b}$, $\Omega_{\rm m}$, $h$ and $n_{\rm s}$. We plot the root mean squared error on $\log F$ as a function of model length for the training and validation sets separately. The model given in \cref{eq:pk_lin_fit} has a model length of 77, as indicated by the dotted line.}
    \label{fig:pk_linear_pareto}
\end{figure}

The root mean squared error for the best function found at each model length is given in \cref{fig:pk_linear_pareto}, where we see that we are able to achieve values of $\mathcal{O}(10^{-3})$ for $\log F$. Unlike for the $\sigma_8$ emulator, we obtain slightly worse losses for the validation set compared to training, however always by less than a factor of two.

Given this set of candidate solutions, we wish to choose one which is sufficiently accurate for current applications yet is sufficiently compact to be interpretable. In \cref{fig:pk_linear_pareto},  one observes a plateau in accuracy between model lengths $\sim65-80$ and thus it seems reasonable to choose a solution in this regime, since doubling the model length only achieves approximately a factor of two improvement in fit beyond this point. Moreover, beyond this point the training and validation curves begin to deviate, suggesting a degree of overfitting.

We choose to report the model of length 77, as indicated by the dotted line in \cref{fig:pk_linear_pareto}, since this provided one of the most interpretable solutions, and achieved a sub-percent error for 95\% ($2\sigma$) of the cosmological parameters considered, for both the training and validation set.
After some simplification, this can be written as

\afterpage{\FloatBarrier} 
\begin{widetext}
    \begin{equation}
    \label{eq:pk_lin_fit}
    \begin{split}
       \log F & \approx  b_0 h - b_{1} \\
        & + \left( \frac{b_2 \Omega_{\rm b}}{\sqrt{h^2 + b_3}} \right)^{b_4 \Omega_{\rm m}}
        \left[
            \frac{b_5 k - \Omega_{\rm b}}{\sqrt{b_6 + (\Omega_{\rm b} - b_7 k)^2}} b_8 (b_9 k)^{-b_{10} k} 
            \cos \left( b_{11} \Omega_{\rm m} - \frac{b_{12} k}{\sqrt{b_{13} + \Omega_{\rm b}^2}} \right)
            - b_{14} \left( \frac{b_{15} k}{\sqrt{1 + b_{16} k^2}} - \Omega_{\rm m} \right) \cos \left( \frac{b_{17} h}{\sqrt{1 + b_{18} k^2}} \right)
        \right] \\
        & + b_{19} (b_{20} \Omega_{\rm m} + b_{21} h - \log(b_{22} k) + (b_{23} k)^{- b_{24} k}) \cos \left( \frac{b_{25}}{\sqrt{1 + b_{26} k^2}} \right) \\
        & + (b_{27} k)^{-b_{28} k} \left( b_{29} k - \frac{b_{30} \log(b_{31} k)}{\sqrt{b_{32} + (\Omega_{\rm m} - b_{33} h)^2}} \right)  \cos \left(  b_{34} \Omega_{\rm m} - \frac{b_{35} k}{\sqrt{b_{36} + \Omega_{\rm b}^2}} \right),
    \end{split}
    \end{equation}
\end{widetext}
\noindent
where the best-fit parameters for this function are given in \cref{tab:pk_linfit_params}. We find that there are 37 different parameters required for this fit, far fewer than would be used if one were to emulate this with a neural network.
We note that, if we used a method based on the ``score'' approach of \textsc{pysr} \citep{cranmer2020discovering,Cranmer_2023} (see \cref{sec:SR}) to choose our model -- defining the loss to be the mean squared error on the training set and computing the derivative with respect to model length -- then we would have chosen the model of length 80. This has an almost identical functional form to \cref{eq:pk_lin_fit}, so our results are not sensitive to the exact model selection method.

\begin{figure*}[htb]
    \centering
    \includegraphics[width=\textwidth]{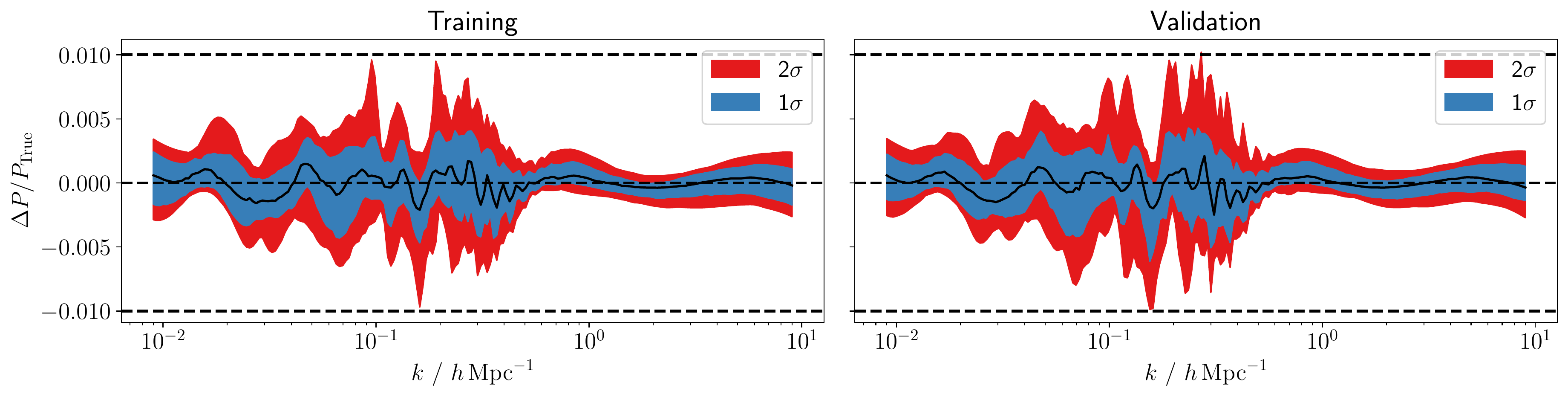}
    \caption{Distribution of fractional errors as a function of $k$ on the linear matter power spectrum across all cosmologies in the training and validation sets, as compared to the predictions of \camb{}. The bands give the 1 and $2\sigma$ values. The dotted line corresponds to a 1\% error, and we see that our expression achieves this for all cosmologies and values of $k$ considered, with a root mean squared fractional error of 0.2\%.}
    \label{fig:residuals_pk_linear}
\end{figure*}

We plot this fit and the residuals compared to \camb{} for the Planck 2018 cosmology in \cref{fig:planck_fit_linear}, which we note is not included in either our training or validation sets. One can see that the difference between the true power spectrum and our analytic fit is almost imperceptible, and in the residuals plot we see that for all $k$ considered, the fractional error does not exceed 0.3\%. This is smaller than the error on $\log F$ given in \cref{fig:pk_linear_pareto}, since we compare at the level of the full $P(k; \bm{\theta})$, such that a moderate error on $\log F$ becomes very small once substituted into \cref{eq:Pk_residual_definition}. This is shown in \cref{fig:residuals_pk_linear}, where we plot the distribution of fractional residuals in $P(k; \bm{\theta})$ for all the cosmologies in the training and validations sets. We obtain sub-percent level predictions for all cosmologies and values of $k$ considered, with a root mean squared fractional error of 0.2\%.

Part of the appeal of a symbolic emulator is the possibility for interpretability and to easily identify what information used in the input is used to make the prediction. To begin, we note that, although we obtained our emulator by varying $\sigma_8$, $\Omega_{\rm b}$, $\Omega_{\rm m}$, $h$ and $n_{\rm s}$, we see that \cref{eq:pk_lin_fit} contains neither $\sigma_8$ nor $n_{\rm s}$. For the linear matter power spectrum, one expects that $A_{\rm s}$ and $n_{\rm s}$ only appear as a multiplicative factor of $A_{\rm s} k^{n_{\rm s} - 1}$, with all other terms independent of these parameters. Given that the \citeauthor{Eisenstein_1998} term already contains this expression, it is unsurprising that $\log F$ is independent of $n_{\rm s}$. Indeed, if it did appear, this would indicate a degree of overfitting. Since $\sigma_8$ is not proportional to $A_{\rm s}$ (see \cref{eq:sigma8_fit}), we cannot use the same argument to explain the lack of its appearance in our expression, but can conclude that a combination of the \citeauthor{Eisenstein_1998} term and the first line of \cref{eq:pk_lin_fit} can sufficiently approximate $A_{\rm s}$, since this line is $k$ independent and thus contributes to an overall offset for the emulator.

\begin{figure}[htb]
    \centering
    \includegraphics[width=\columnwidth]{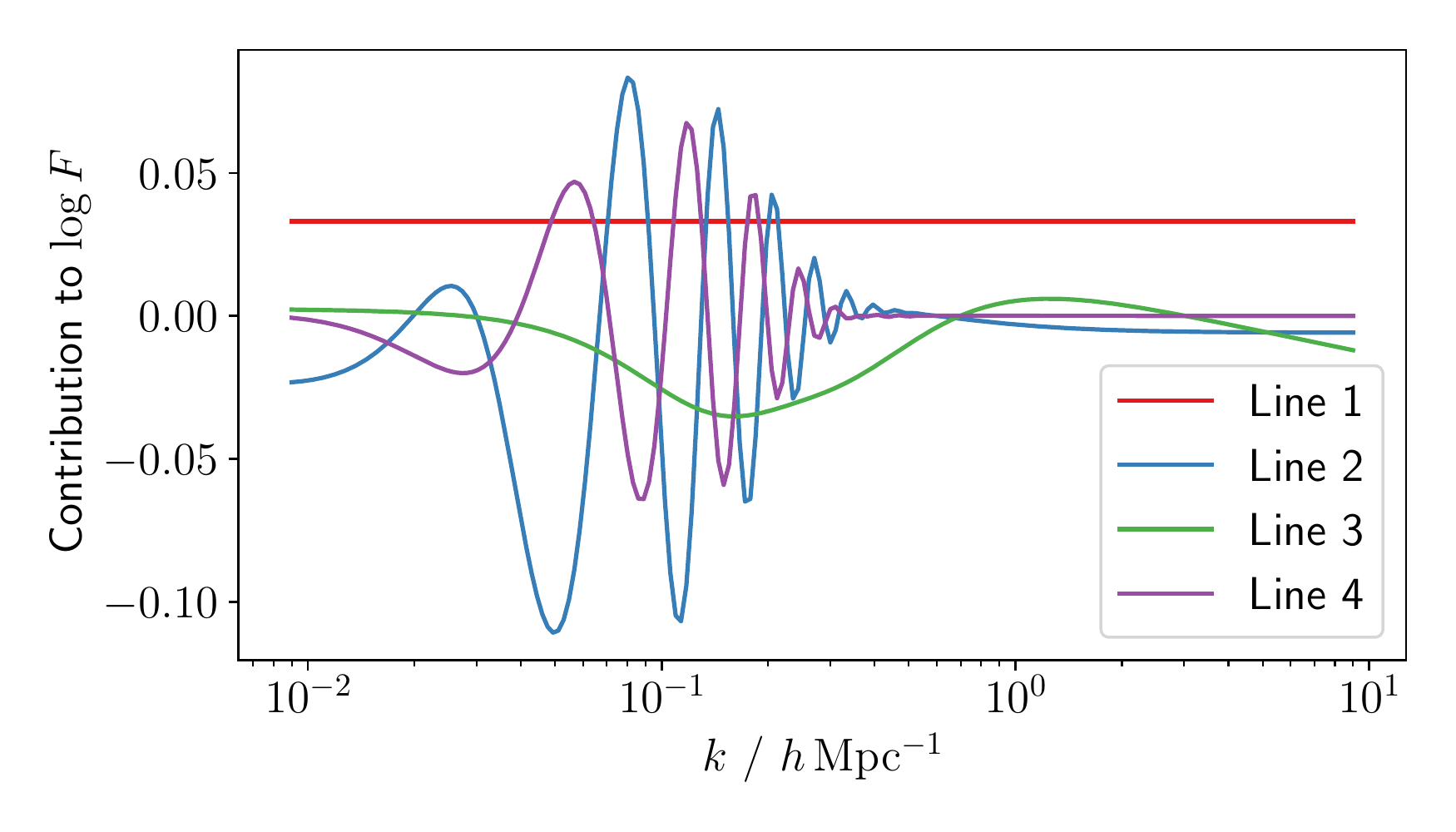}
    \caption{Contributions to $\log F$ from our emulator as a function of $k$ for the Planck 2018 cosmology. The line numbers indicated in the legend correspond to the line in \cref{eq:pk_lin_fit}. One sees that the first term provides an overall offset, the second and fourth capture the BAO signal, and the third term contains a broad oscillation and then matches on to the decaying residual at high $k$.}
    \label{fig:logF_terms}
\end{figure}

Turning to the remaining lines of \cref{eq:pk_lin_fit}, we observe that each term contains an oscillation modulated by a $k$- and cosmology-dependent damping. Despite there being four such terms across the remaining three lines, we find that we can split these into two pairs with the same structure of the oscillations. Firstly, we have cosines with an argument proportional to $1 / \sqrt{1 + b k^2}$, for some constant $b$. This functional form ($x/\sqrt{1 + y^2}$) arises due to the inclusion of the analytic quotient operator, which also explains why the constant $1$ appears multiple times in \cref{eq:pk_lin_fit}.
These terms give oscillations which vary slowly as a function of $k$. In particular, as plotted in \cref{fig:logF_terms}, the third line of \cref{eq:pk_lin_fit} contains approximately one cycle of oscillation across the range of $k$ considered, with a minimum during the BAO part of the power spectrum, and a maximum just afterwards. Beyond this point, this term fits the non-oscillatory, decaying part of the residual beyond $k\sim 1 \, h {\rm \, Mpc}^{-1} $ (compare the middle panel of \cref{fig:planck_fit_linear} to the third term plotted in \cref{fig:logF_terms}).

The remaining oscillatory terms are of the form $\cos(\omega k + \phi)$. The phase, $\phi$, of these oscillations is proportional to the total matter density, $\Omega_{\rm m}$, such that changing this parameter at fixed $\Omega_{\rm b}$ shifts the BAOs to peak at different values of $k$. The frequency of these oscillations is $\omega \propto 1 / \sqrt{b + \Omega_{\rm b}^2}$ for some parameter $b$, such that cosmologies with a higher fraction of baryons have many more cycles of BAO in a given range of $k$, as one would physically expect. 
From \cref{fig:logF_terms}, one can see how the second and fourth lines of \cref{eq:pk_lin_fit} capture the BAO signal with opposite signs, such that they combine to give the familiar damped oscillatory feature. 
Using $\Omega_{\rm b}h^2 = 0.02242$ and $h  = 0.6766$, as appropriate for the Planck 2018 cosmology \citep{Planck_VI_2018}, the frequency of the oscillations are $b_{12} / \left( h \sqrt{b_{13} + \Omega_{\rm b}^2} \right) = 146.5 {\rm \, Mpc}$ and $b_{35} / \left(h \sqrt{b_{36} + \Omega_{\rm b}^2}\right) = 145.8 {\rm \, Mpc}$, and are thus approximately equal to the sound horizon, which is $r_\ast = 144.6 {\rm \, Mpc}$ for this cosmology.
One can therefore view these frequencies as symbolic approximations to the sound horizon, although we refer the reader to \citet{Aizpuru_2021} for alternative SR fits.

Thus, although we did not enforce physically motivated terms in the equation search, we see that simple oscillatory contributions for the BAOs have emerged and thus our symbolic emulator is not merely a high order series expansion, but contains terms which are both compact and interpretable. We find that such terms exist in many functions given in \cref{fig:pk_linear_pareto}, however we find that using shorter run times for \operon{} of only 2-4 hours (compared to approximately 24 hours on a single node of 128 cores for our fiducial analysis) do not provide as interpretable expressions as \cref{eq:pk_lin_fit}.

\begin{table}[]
    \caption{Best-fit parameters for the linear matter power spectrum emulator given in \cref{eq:pk_lin_fit}.}
    \centering
    \begin{tabular}{c|l|c|l}
    Parameter & Value & Parameter & Value \\
    \hline\hline
    $b_{0}$ & 0.0545& $b_{19}$ & 0.0111\\
$b_{1}$ & 0.0038& $b_{20}$ & 5.35\\
$b_{2}$ & 0.0397& $b_{21}$ & 6.421\\
$b_{3}$ & 0.1277& $b_{22}$ & 134.309\\
$b_{4}$ & 1.35& $b_{23}$ & 5.324\\
$b_{5}$ & 4.0535& $b_{24}$ & 21.532\\
$b_{6}$ & 0.0008& $b_{25}$ & 4.742\\
$b_{7}$ & 1.8852& $b_{26}$ & 16.6872\\
$b_{8}$ & 0.1142& $b_{27}$ & 3.078\\
$b_{9}$ & 3.798& $b_{28}$ & 16.987\\
$b_{10}$ & 14.909& $b_{29}$ & 0.0588\\
$b_{11}$ & 5.56& $b_{30}$ & 0.0007\\
$b_{12}$ & 15.8274& $b_{31}$ & 195.498\\
$b_{13}$ & 0.0231& $b_{32}$ & 0.0038\\
$b_{14}$ & 0.8653& $b_{33}$ & 0.2767\\
$b_{15}$ & 0.8425& $b_{34}$ & 7.385\\
$b_{16}$ & 4.554& $b_{35}$ & 12.3961\\
$b_{17}$ & 5.117& $b_{36}$ & 0.0134\\
$b_{18}$ & 70.0234& & \\
    \end{tabular}
    \tablefoot{Although units are excluded in this table, the units for each parameter are easily obtained by noting that these are defined assuming that $k$ is measured in $h\, {\rm Mpc^{-1}}$ in \cref{eq:pk_lin_fit}.}
    \label{tab:pk_linfit_params}
\end{table}

As a note of caution, one can identify a few terms in \cref{eq:pk_lin_fit} which will become problematic if extrapolated to values of $k$ much smaller than those used to train the emulator, namely those containing $\log(k)$ and $k$ raised to a power proportional to $k$. For $k \lesssim 10^{-3} \, h {\rm \, Mpc}^{-1}$ this can lead to an error on $P(k)$ of more than one percent. Although one is likely cosmic-variance dominated in this regime so such errors should not be problematic, we know that the \citeauthor{Eisenstein_1998} provides a very good approximation, and thus we suggest that \cref{eq:pk_lin_fit} is included in a piece-wise fit, such that it is only used approximately in the range of $k$ which were used to obtain it. A similar effect is seen if one extrapolates to higher $k$ than considered here. Although this is far beyond the validity of the linear approximation, we caution that applying any parameterisation of the non-linear power spectrum that depends on the linear one may suffer from potentially catastrophic extrapolation failures at high $k$ if used beyond the $k$ range considered here. Again, it is potentially advisable to just use the \citeauthor{Eisenstein_1998} fit in this regime.

In light of the potential for significant efficiency improvements in cosmological analyses through the use of symbolic approximations, it is informative to compare the run times of our approach to the standard computation of the linear matter power spectrum. To do this, we evaluated the redshift-zero linear matter power spectrum 1000 times on an 
Intel Xeon E5-4650 CPU 
at the Planck 2018 cosmology \citep{Planck_VI_2018} using \camb{} \citep{Lewis_2000}, the \bacco{} \citep{Angulo_2021} neural network emulator and using our formulae, where we considered both a \python{} and \fortran{} implementation, demonstrating the ease at which one can change programming language when using symbolic emulators.
We found that \camb{} takes an average of $0.18{\rm \, s}$ to evaluate $P(k)$, which is significantly slower than the \bacco{} emulator, which requires just $6.9{\rm \, ms}$. However, our approach is even faster, requiring just $850 {\rm \, \mu s}$ when written in \python{} and $190 {\rm \, \mu s}$ in \fortran.
This is approximately 950 times faster than \camb{} and 36 times faster than \bacco.

If the reader wishes to use a more accurate, yet less interpretable, emulator, we provide the most accurate equation found in \cref{app:most_accurate}, which has a model length of 142, with 73 parameters and yields a root mean of squared fractional errors on $P(k)$ of 0.1\% for both the training and validation sets.

\section{Discussion and conclusion}
\label{sec:Conclusion}

In this paper we have found analytic approximations to $\sigma_8$ (\cref{eq:sigma8_fit}) and the linear matter power spectrum (\cref{eq:pk_lin_fit}) as a function of cosmological parameters which are accurate to sub-percent levels. 
In the case of $\sigma_8$, the simple yet accurate expression we have identified can be easily inverted to obtain $A_{\rm s}$ as a function of $\sigma_8$ and the other cosmological parameters. 
Our approximation to $P(k)$ is built by fitting the residuals between the output of a Boltzmann solved (\camb) and the physics-inspired approximation of \citet{Eisenstein_1998}. 
As such, unlike neural network or Gaussian process based approaches, our expression explicitly captures many physical processes (and is thus interpretable) whilst still achieving sub-percent accuracy.

This work is the first step in a programme of work dedicated to obtaining analytic approximations to $P(k)$ which can be used in current and future cosmological analyses.
In this paper we have focused on the linear $P(k)$, i.e. the power spectrum of the linearly evolved density fluctuations.
Although this approach is valid on large scales, the real Universe is non-linear, such that non-linear corrections are required at $k\gtrsim 10^{-1} \, h {\rm \, Mpc}^{-1}$ to accurately model the observed matter power spectrum across a wider range of scales.
In \citet{Bartlett_2024} we extend
our framework to capture such non-linear physics and to include redshift dependence in our emulator.
Finally, in our emulator we have considered a $\Lambda$CDM Universe with massless neutrinos.
In the future we will add corrections to the expressions found in this work to incorporate the effects of massive neutrinos and include beyond $\Lambda$CDM effects, such as a $w_0-w_a$ parametrisation of dark energy.

We have demonstrated that, despite the temptation to blindly apply black-box methods such as neural networks to approximate physically useful functions, even in ostensibly challenging situations such as the matter power spectrum, one can achieve the required precision with relatively simple analytic fits. Given the unknown lifetime of current codes upon which numerical ML approximations are built and the ease of copying a few mathematical functions into your favourite programming language, finding analytic expressions allows one to more easily future-proof such emulators and should therefore be encouraged wherever possible.

\section*{Acknowledgements}

We thank Bartolomeo Fiorini for useful comments and suggestions.
DJB is supported by the Simons Collaboration on ``Learning the Universe.''
LK was supported by a Balzan Fellowship.
HD is supported by a Royal Society University Research Fellowship (grant no. 211046).
PGF acknowledges support from STFC and the Beecroft Trust. 
BDW acknowledges support from the Simons Foundation.
DA acknowledges support from the Beecroft Trust, and from the Science and Technology Facilities Council through an Ernest Rutherford Fellowship, grant reference ST/P004474/1. MZ is supported by STFC.
We made extensive use of computational resources at the University of Oxford Department of Physics, funded by the John Fell Oxford University Press Research Fund, and at the Institut d'Astrophysique de Paris.

For the purposes of open access, the authors have applied a Creative Commons Attribution (CC BY) licence to any Author Accepted Manuscript version arising.

The data underlying this article will be shared on reasonable request to the corresponding author.
We provide 
\python{} and \fortran{}
implementations of \cref{eq:sigma8_fit,eq:Pk_residual_definition,eq:pk_lin_fit,eq:pk_most_accurate} at \url{https://github.com/DeaglanBartlett/symbolic_pofk}.

\bibliographystyle{aa} 
\bibliography{references} 

\begin{appendix} 
\onecolumn
\section{Most accurate analytic expression found for linear power spectrum}
\label{app:most_accurate}

The expression we report for an analytic approximation for the linear matter power spectrum (\cref{eq:pk_lin_fit}) is not the most accurate one found, but the one which we deemed to appropriately balance accuracy, simplicity, and interpretability. 
It may be desirable to have a more accurate symbolic expression if interpretability is not a concern. In this case one may wish to use the most accurate equation found, which is
\begin{equation}
    \label{eq:pk_most_accurate}
    \begin{split}
        100 \log F &\approx
        c_{0} k 
        + c_{1} \left(\Omega_{\rm b} c_{2} - \frac{c_{3} k}{\sqrt{c_{4} + k^{2}}}\right) \left(\frac{c_{34} \left(c_{35} k\right)^{- c_{36} k}}{\sqrt{c_{39} + \left(- \Omega_{\rm b} + \Omega_{\rm m} c_{37} - c_{38} h\right)^{2}}} - \cos{\left(\Omega_{\rm m} c_{32} - c_{33} k \right)}\right) \\
        & \times \Bigg(\frac{c_{17} \left(c_{25} k\right)^{- c_{26} k} \left(\left(\Omega_{\rm b} c_{18} + \Omega_{\rm m} c_{19} - c_{20} h\right) \cos{\left(\Omega_{\rm m} c_{21} - c_{22} k \right)} + \cos{\left(c_{23} k - c_{24} \right)}\right)}{\sqrt{c_{31} + \left(\frac{c_{27} \left(- \Omega_{\rm m} c_{28} + c_{29} k\right)}{\sqrt{c_{30} + k^{2}}} - k\right)^{2}}} \\
        & - \frac{c_{5} \left(\Omega_{\rm m} c_{12} + c_{13} k\right)^{- c_{14} k} \left(\Omega_{\rm m} c_{6} - c_{7} k + \left(\Omega_{\rm b} c_{8} - c_{9} k\right) \cos{\left(\Omega_{\rm m} c_{10} - c_{11} k \right)}\right)}{\sqrt{c_{16} + \left(\Omega_{\rm b} c_{15} + k\right)^{2}}}\Bigg) \\
        & - c_{40} \left(\Omega_{\rm m} c_{41} - c_{42} h + c_{43} k + \frac{c_{44} k}{\sqrt{c_{45} + k^{2}} \sqrt{c_{47} + \left(- \Omega_{\rm m} - c_{46} h\right)^{2}}} - \frac{c_{48} \left(\Omega_{\rm m} c_{49} + c_{50} k\right)}{\sqrt{c_{51} + k^{2}}}\right) \cos{\left(\frac{c_{52} k}{\sqrt{c_{53} + k^{2}} \sqrt{c_{55} + \left(\Omega_{\rm m} c_{54} - k\right)^{2}}} \right)} \\
        & - c_{56} - \frac{c_{57} \left(\Omega_{\rm m} c_{67} + c_{68} k\right)^{- c_{69} k} \left(\Omega_{\rm m} c_{58} - c_{59} k + \left(- \Omega_{\rm b} c_{60} - \Omega_{\rm m} c_{61} + c_{62} h\right) \cos{\left(\Omega_{\rm m} c_{63} - c_{64} k \right)} + \cos{\left(c_{65} k - c_{66} \right)}\right)}{\sqrt{\frac{c_{70} \left(\Omega_{\rm b} + \frac{c_{71} h}{\left(c_{72} + k^{2}\right)^{0.5}}\right)^{2}}{c_{73} + k^{2}} + 1.0}}
        .
    \end{split}
\end{equation}
This equation has 73 parameters, which is approximately twice as many as \cref{eq:pk_lin_fit}, yet one only gains a factor of two in the fractional root mean squared error. The best-fit parameter values are reported in \cref{tab:pk_best_params}. We note that this function is the direct output of \operon{} and is thus over-parameterised so that some simplification could be applied. For example, one only needs two of $c_1$, $c_2$ and $c_3$ as these only appear as $c_1c_2$ and $c_1c_3$. Since we only provide this expression as a precise emulator and do not attempt to interpret its terms, we choose not to apply any simplifications (although see \citet{deFranca_2023} for an automated method to do this).

\begin{table*}[]
    \caption{Best-fit parameters for the most accurate linear matter power spectrum emulator found, reported in \cref{app:most_accurate}.}
    \centering
    \begin{tabular}{c|l|c|l|c|l|c|l}
    Parameter & Value & Parameter & Value & Parameter & Value & Parameter & Value \\
    \hline\hline
$c_{0}$ & 5.1439& $c_{19}$ & 19.855& $c_{38}$ & 0.0177& $c_{57}$ & 0.867\\
$c_{1}$ & 0.867& $c_{20}$ & 15.939& $c_{39}$ & 0.0146& $c_{58}$ & 2.618\\
$c_{2}$ & 8.52& $c_{21}$ & 9.547& $c_{40}$ & 0.867& $c_{59}$ & 2.1\\
$c_{3}$ & 0.2920& $c_{22}$ & 97.34& $c_{41}$ & 32.371& $c_{60}$ & 114.391\\
$c_{4}$ & 0.0310& $c_{23}$ & 94.83& $c_{42}$ & 7.058& $c_{61}$ & 13.968\\
$c_{5}$ & 0.0033& $c_{24}$ & 1.881& $c_{43}$ & 6.075& $c_{62}$ & 11.133\\
$c_{6}$ & 240.234& $c_{25}$ & 3.945& $c_{44}$ & 16.311& $c_{63}$ & 4.205\\
$c_{7}$ & 682.449& $c_{26}$ & 11.151& $c_{45}$ & 0.0025& $c_{64}$ & 100.376\\
$c_{8}$ & 2061.023& $c_{27}$ & 0.0004& $c_{46}$ & 0.1632& $c_{65}$ & 106.993\\
$c_{9}$ & 6769.493& $c_{28}$ & 26.822& $c_{47}$ & 0.0771& $c_{66}$ & 3.359\\
$c_{10}$ & 7.125& $c_{29}$ & 230.12& $c_{48}$ & 0.0522& $c_{67}$ & 1.539\\
$c_{11}$ & 108.136& $c_{30}$ & 0.0009& $c_{49}$ & 22.722& $c_{68}$ & 1.773\\
$c_{12}$ & 6.2& $c_{31}$ & 1.0796$\times10^{-5}$& $c_{50}$ & 774.688& $c_{69}$ & 18.983\\
$c_{13}$ & 2.882& $c_{32}$ & 3.162& $c_{51}$ & 0.0027& $c_{70}$ & 0.3838\\
$c_{14}$ & 59.585& $c_{33}$ & 99.918& $c_{52}$ & 1.0337& $c_{71}$ & 0.0024\\
$c_{15}$ & 0.1384& $c_{34}$ & 0.1210& $c_{53}$ & 0.0058& $c_{72}$ & 1.2865$\times10^{-7}$\\
$c_{16}$ & 1.0825$\times10^{-5}$& $c_{35}$ & 0.495& $c_{54}$ & 0.2472& $c_{73}$ & 2.0482$\times10^{-8}$\\
$c_{17}$ & 0.0033& $c_{36}$ & 12.091& $c_{55}$ & 0.29& & \\
$c_{18}$ & 85.791& $c_{37}$ & 0.6070& $c_{56}$ & 0.241& & \\
    \end{tabular}
    \tablefoot{Although units are excluded in this table, the units for each parameter are easily obtained by noting that these are defined assuming that $k$ is measured in $h\, {\rm Mpc^{-1}}$.}
    \label{tab:pk_best_params}
\end{table*}

\end{appendix}

\end{document}